\documentclass[%
,twocolumn%
,secnumarabic%
,amssymb,aps,pra,nobibnotes]{revtex4-1}
\usepackage{graphicx}
\usepackage{amsmath}
\expandafter\ifx\csname package@font\endcsname\relax\else
 \expandafter\expandafter
 \expandafter\usepackage
 \expandafter\expandafter
 \expandafter{\csname package@font\endcsname}%
\fi
\DeclareRobustCommand\substyle{\name@idx{document substyle}}%
\DeclareRobustCommand\classoption{\name@idx{document class option}}%
\DeclareRobustCommand\classname{\name@idx{document class}}%
\def\name@idx#1#2{%
 {\ttfamily#2}%
 \index{#2\space#1=\string\ttt{#2}\space#1}\index{#1>#2=\string\ttt{#2}}%
}%

\begin{document}

\title{Continuous-variable measurement-device-independent quantum key distribution with virtual photon subtraction}

\author{Yijia Zhao$^1$}
\author{Yichen Zhang$^{1,2,}$}
\email{zhangyc@bupt.edu.cn}
\author{Bingjie Xu$^3$}
\author{Song Yu$^{1,}$}
\email{yusong@bupt.edu.cn}
\author{Hong Guo$^{2}$}

\address{$^1$State Key Laboratory of Information Photonics and Optical Communications, Beijing University of Posts and Telecommunications, Beijing 100876, China}
\address{$^2$State Key Laboratory of Advanced Optical Communication System and Network, School of Electronics Engineering and Computer Science and Center for Quantum Information Technology, Peking University, Beijing 100871, China}
\address{$^3$Science and Technology on Security Communication Laboratory, Institute of Southwestern Communication, Chengdu 610041, China}

\begin{abstract}
The method of improving the performance of continuous-variable quantum key distribution protocols by post-selection has been recently proposed and verified. In continuous-variable measurement-device-independent quantum key distribution (CV-MDI QKD) protocols, the measurement results are obtained from untrusted third party Charlie. There is still not an effective method of improving CV-MDI QKD by the post-selection with untrusted measurement. We propose a method to improve the performance of coherent-state CV-MDI QKD protocol by virtual photon subtraction via non-Gaussian post-selection. The non-Gaussian post-selection of transmitted data is equivalent to an ideal photon subtraction on the two-mode squeezed vacuum state, which is favorable to enhance the performance of CV-MDI QKD. In CV-MDI QKD protocol with non-Gaussian post-selection, two users select their own data independently. We demonstrate that the optimal performance of the renovated CV-MDI QKD protocol is obtained with the transmitted data only selected by Alice. By setting appropriate parameters of the virtual photon subtraction, the secret key rate and tolerable excess noise are both improved at long transmission distance. The method provides an effective optimization scheme for the application of CV-MDI QKD protocols.
\begin{description}
\item[PACS numbers]
03.67.Dd, 03.67.Hk
\end{description}
\end{abstract}

\pacs{Valid PACS appear here}
\maketitle


\section{Introduction}

Continuous-variable quantum key distribution (CV-QKD) allows two users to achieve unconditional secure key distribution~\cite{CV3,CV4}, which is based on the basic principles of quantum mechanics. CV-QKD protocol based on Gaussian modulated coherent states (GMCS)~\cite{GMCS1,GMCS2} has unique potentials in the high secret key rate and the compatibility with the commercial communication system. The experimental research of CV-QKD has gradually matured in recent years~\cite{EXPERIMENT1,EXPERIMENT2}. Recently, the field tests of the system over commercial fibers has exceeded 50km and obtained the same order secret key rate as the DV system~\cite{EXPERIMENT4}, where the system can support the practical application within metropolitan area. The GMSC CV-QKD protocols have been proved to be unconditionally secure in theory~\cite{theorysecurity1,theorysecurity2,theorysecurity3,theorysecurity4,theorysecurity5} under some ideal assumptions. However, the practical security loopholes which arise from the imperfect devices are still major obstacle to the development of CV-QKD.

Recently, several quantum attack strategies against the detection were proposed such as local oscillator intensity and calibration attack ~\cite{CVattack1,CVattack2,CVattack3,CVattack4}. Obviously, each known loopholes of practical CV-QKD system can be defended by corresponding countermeasure, which will increase the complexity and reduce the reliability of the system. Moreover, it is difficult to find an effective way to prevent an unknown attack. The continuous-variable measurement-device-independent quantum key distribution (CV-MDI QKD) protocols~\cite{CVMDI1,CVMDI2} are proposed to counter general attacks against detection loopholes in practical systems~\cite{CVMDI3,CVMDI4,CVMDI5,CVMDI6,CVMDI7,CVMDI8,CVMDI9}. In CV-MDI QKD protocols, the secret key between two legal parties is established by the measurement results of an untrusted third party. On the one hand, CV-MDI QKD protocols are immune to all attacks on detection and therefore have higher practical security. On the other hand, the methods~\cite{NLAPS1,NLACV1,NLACV2} that optimize CV-QKD protocols at the detection side are no longer applicable in the CV-MDI QKD protocols duo to the untrusted measurement results makes it difficult to optimize CV-MDI protocols at the detection side.

The non-Gaussian post-selection of transmitted data which is equivalent to the virtual photon subtraction of coherent state source can enhance the performance of CV-QKD protocols~\cite{PSEB,PSPS,PSPSTWO}. Thus, non-Gaussian post-selection is a superior way to improve the performance of the CV-MDI QKD protocols in practice. In this paper, we propose a method to improve the performance of coherent-state CV-MDI QKD protocols by virtual photon subtraction. Non-Gaussian post-selection is applied on the transmitted data so that the protocols is modified without using the untrusted measurement results. We demonstrate that the non-Gaussian post-selection will not benefit in any case. The performance of renovated protocol is only improved in the case where Alice uses the reasonable parameters. Moreover, Eve could apply two independent attacks or correlated attacks on two quantum channels. The numerical simulation results show that the improvement of virtual photon subtraction is still effective under the optimal correlated attacks~\cite{CVMDI2,CVMDI4}.

This paper is organized as follows: In Sec.~\ref{sec:2}, we propose the CV-MDI QKD protocol with virtual photon subtraction and derive the secret key rate of the renovated protocol. In Sec.~\ref{sec:3}, we present the performance of the CV-MDI QKD protocol with virtual photon subtraction through numerical simulation and the optimal scheme of the renovated protocol. Conclusions are given In Sec.~\ref{sec:4}.

\section{\label{sec:2}CV-MDI QKD protocol with virtual photon subtraction}

In this section, we first introduce the CV-MDI QKD with virtual photon subtraction. Then we present the security analysis of the renovated protocol where Eve implements correlated attacks on two quantum channels. The optimal correlated attacks have been demonstrated to be more effective to CV-MDI protocols~\cite{CVMDI4,CVMDI2}, which is more reasonable to evaluate the performance of the protocol.

\subsection*{A. Non-Gaussian post-selection in CV-MDI QKD }

\begin{figure*}[t]
\centering
\includegraphics[width=7in]{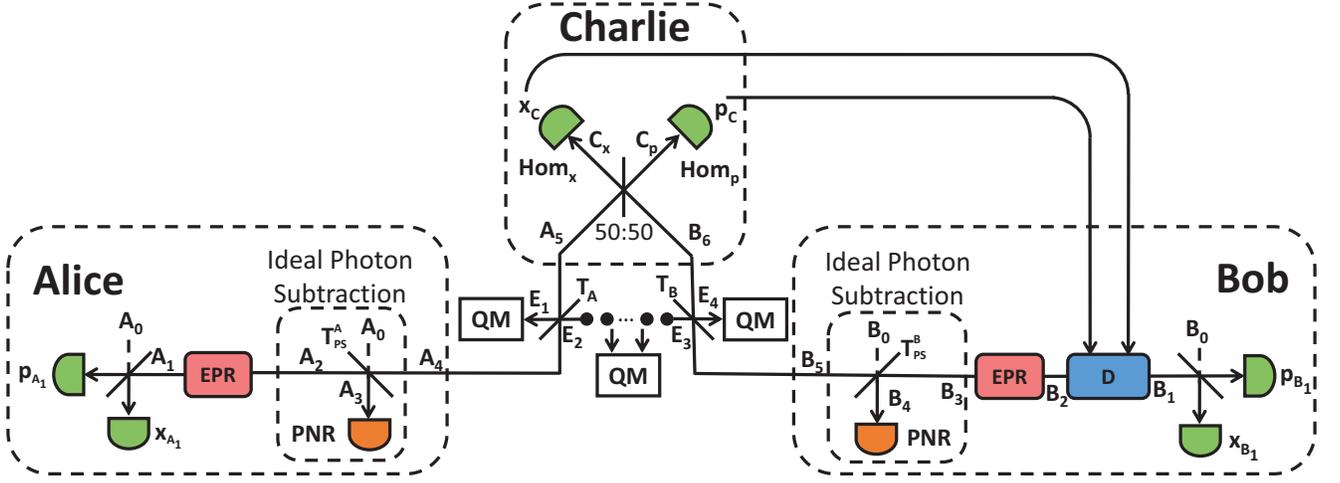}
\caption{ (Color online) Entanglement-based scheme of CV-MDI QKD with virtual photon subtraction. EPR is the two-mode squeezed vacuum state. All green detectors represent homodyne detection. PNR is the photon number resolving detector. $T_A$ ($T_B$) is the channel transmittance from Alice (Bob) to Charlie. D is the displacement operation of Bob. QM is the quantum memory.}
\label{fig1}
\end{figure*}

Although most of the implementation of CV-MDI QKD protocols is based on prepare-and-measure (PM) scheme, the security of the protocol with non-Gaussian post-selection still needs to build a entanglement-based (EB) model. It has been confirmed that the preparation-measurement (PM) scheme of non-Gaussian post-selection is equivalent to the entanglement-based scheme of realistic photon subtraction with an ideal photon number resolving detector~\cite{PSPS,PNR}. In order to make the key parameters of the post-selection easier to understand, we show the EB scheme which is easier to security analysis in Fig.~\ref{fig1}.

In the EB scheme, the coherent state is prepared by applying heterodyne detection on one mode of the two-mode squeezed vacuum (TMSV) state. We will show the equivalence between non-Gaussian post-selection and photon subtraction on one mode of her TMSV state~\cite{PSPS}. The mode ${A_4}$ with $k$ photons subtraction is described by

\begin{equation}
\begin{array}{l}
\rho _{{A_4}}^k = \int {\int {\frac{{P_{PS}^k\left( {k\left| {{x_{A_1}},{p_{A_1}}} \right.} \right)}}{{P_{PS}^k}}} } {P_{{x_{A_1}}{p_{A_1}}}}\\
{\kern 1pt} {\kern 1pt} {\kern 1pt} {\kern 1pt} {\kern 1pt} {\kern 1pt} {\kern 1pt} {\kern 1pt} {\kern 1pt} {\kern 1pt} {\kern 1pt} {\kern 1pt} {\kern 1pt} {\kern 1pt} {\kern 1pt} {\kern 1pt} {\kern 1pt} {\kern 1pt} {\kern 1pt} d{x_{A_1}}d{p_{A_1}}\left| {\sqrt {{T_{PS}}} \alpha } \right\rangle \langle \sqrt {{T_{PS}}} \alpha |,
\end{array}
\label{NG}
\end{equation}
and
\begin{equation}
\begin{array}{l}
P_{PS}^k\left( {k\left| {{x_{A_1}},{p_{A_1}}} \right.} \right) = {\left| {\left\langle {k\left| {\sqrt {1 - {T_{PS}}} \alpha } \right\rangle } \right.} \right|^2},
\end{array}
\end{equation}
where $\alpha  = \sqrt 2 \lambda ({x_{A_1}} + i{p_{A_1}})$, $\lambda  = \sqrt {\frac{{V - 1}}{{V + 1}}} $, $V$ is the variance of the TMSV
state, ${T_{PS}}$ is transmittance of photon subtraction, $k$ is the number of photon subtraction, $P_{PS}^k\left( {k\left| {{x_{A_1}},{p_{A_1}} }\right.} \right)$ is the success probability of k-photon subtraction with Alice's heterodyne measurement result $\left( {{x_{A_1}},{p_{A_1}}} \right)$ and ${P_{{x_{A_1}}{p_{A_1}}}}$ is the two-dimensional Gaussian distribution of Alice¡¯s heterodyne measurement result $\left( {{x_{A_1}},{p_{A_1}}} \right)$.

Then we show the case where photon subtraction is not implemented ($T_{PS}=1$), the ${A_4}$ mode is described by

\begin{equation}
{\rho _{{A_4}}} = \int {\int {{P_{{x_{A_1}}{p_{A_1}}}}d{x_{A_1}}d{p_{A_1}}\left| \alpha  \right\rangle \left\langle  \alpha  \right|} }.
\end{equation}
By Comparing ${\rho _{{A_4}}}$ with $\rho _{{A_4}}^k$ , the photon subtracted TMSV state is achieved by selecting Alice's heterodyne measurement result $\left( {{x_{A_1}},{p_{A_1}}} \right)$ with a probability ${P_s} = P_{PS}^k\left( {k\left| {{x_{A_1}},{p_{A_1}}} \right.} \right)$. According to the equivalence relation between Alice's heterodyne measurement results in the EB scheme and Gaussian modulation data in the PM scheme, the corresponding non-Gaussian post-selection probability of each pair modulation data is derived.

CV-MDI QKD system contains two sources located at Alice and Bob's sides respectively, and the influence of each side's non-Gaussian post-selection should be analyzed independently. To know the influence of one-side operation on the other side's data, we first assume the non-Gaussian post-selection is only implemented by Alice. Alice selects her data $\left( {{x_A},{p_A}} \right)$ with a selection probability $P_S^A = P_{PS}^k\left( {k\left| {{x_A},{p_A}} \right.} \right)$ and the data of Bob is also selected by the probability $P_S^A$. Then the original Gaussian state of Alice will be transformed into a non-Gaussian state as Eq.~\ref{NG}. Since there is no correlation between Alice and Bob's data, the selection probability $P_S^A$ which only depends on $\left( {{x_A},{p_A}} \right)$ is independent on Bob's data. When Bob select his data with selection probability $P_S^A$ as if he select a group of Gaussian data randomly. As a consequence, the original TMSV state will not change. It means that Alice or Bob can select her or his transmitted data independently without change of the other one's. As a result, there are three CV-MDI QKD schemes with non-Gaussian post-selection. We introduce steps of the CV-MDI QKD with non-Gaussian post-selection based on PM scheme as follows and the equivalent EB scheme is shown in Fig~\ref{fig1}:

Step1: Alice and Bob randomly select Gaussian-distribution data pair $\left\{ {{x_A},{p_A}} \right\}$ and $\left\{ {{x_{B_1}},{p_{B_1}}} \right\}$  with variance ${V_{A}-1}$ and ${V_{B}-1}$, respectively. Then they use these data to modulate $x$ quadrature and $p$ quadrature of their coherent state. The Gaussian-modulated coherent states are send to Charlie through the quantum channels controlled by Eve.

Step2: Charlie receives the two modes and input the two modes through a abeam splitter (50:50). He applies homodyne detections on the two output modes and gets the measurement results of $x$ quadrature and $p$ quadrature $\left\{ {{x_C},{p_C}} \right\}$ ,which are announced to Alice and Bob through classical channels.

Step3: Alice keeps her data unchanged. Bob uses ${x_B} = {x_{B_1}} + \mu {x_{C}}$ and ${p_B} = {p_{B_1}} - \mu {p_{C}}$ to modify his data, where $\mu$ is the parameter used to optimize Bob's estimator. This process is similar to the approach in the two-way CV-QKD protocols~\cite{twoway1,twoway2,twoway3,twoway4,twowayOA,twowayS}.

Step4: Before the conventional post processing, Alice and Bob select a part of data with  probability $P_S^A$ and $P_S^B$ respectively(when Alice or Bob does not use non-Gaussian post-selection, $P_S^A$ or $P_S^B$ is set to 100\%), and reveal the selection results. And they should keep the data which is selected by both of them. These data are used to implement the reconciliation and privacy amplification through the classical channel.

Because the security analysis of CV-MDI QKD protocol based on GMCS has been demonstrated to be the well-known one-way CV QKD protocol using coherent states and heterodyne detection~\cite{CVMDI1}, the optimality of Gaussian attack~\cite{theorysecurity1,theorysecurity2} is still applicable in this case. We will lower-bound the secret key rate when we regard the quantum state finally shared by Alice and Bob as a Gaussian state. Although, the original Gaussian-modulated coherent state will be transformed into a non-Gaussian state after non-Gaussian post-selection, the quantum states ${\rho _{A_1}{A_4}}$ and ${\rho _{B_2}{B_5}}$ should still be used as Gaussian states which have the same covariance matrix to estimate the lower bound of secret key rate of the renovated protocol.

\section*{B. Secret key rate of CV-MDI QKD with virtual photon subtraction}


We have introduced that the EB scheme with virtual photon-subtraction and PM scheme with non-Gaussian selection are equivalent. We use the EB scheme which is shown in Fig~\ref{fig1} to derive secret key rate. Although there are three EB schemes of CV-MDI QKD with non-Gaussian selection, the security analysis of three cases can be drawn through an unified form of Alice and Bob's covariance matrix. The covariance matrixes of their sources can be described by

\begin{equation}\label{cmA}
{\gamma _{{A_1}{A_4}}} = \left[ {\begin{array}{*{20}{c}}
{{V_{{A_1}}}{\rm I}}&{{C_{{A_1}{A_4}}}{\sigma _z}}\\
{{C_{{A_1}{A_4}}}{\sigma _z}}&{{V_{{A_4}}}{\rm I}}
\end{array}} \right],
\end{equation}
and
\begin{equation}\label{cmB}
{\gamma _{{B_2}{B_5}}} = \left[ {\begin{array}{*{20}{c}}
{{V_{{B_2}}}{\rm I}}&{{C_{{B_2}{B_5}}}{\sigma _z}}\\
{{C_{{B_2}{B_5}}}{\sigma _z}}&{{V_{{B_5}}}{\rm I}}
\end{array}} \right],
\end{equation}
where ${\rm I}$ is two-dimensional identity matrix and ${\sigma _z} = \left[ {\begin{array}{*{20}{c}}
1&0\\
0&{ - 1}
\end{array}} \right]$.

The elements of Alice's covariance matrix are~\cite{PSPS}

\begin{equation}
\begin{array}{l}
{V_{{A_1}}} = 2{V'_A} - 1,\\
{C_{{A_1}{A_4}}} = 2\sqrt {{T_{PS}}} {\lambda _A}{V'_A},\\
{V_{{A_4}}} = 2{T_{PS}}\lambda _A^2{V'_A} + 1,\\
{V'_A} = \frac{{k + 1}}{{1 - {T_{PS}}\lambda _A^2}}.
\end{array}
\label{pscm}
\end{equation}

The way of deriving the elements of Bob's covariance matrix is same to the Eq.~(\ref{pscm}). All the cases of the modified protocol can be obtained by setting corresponding parameters (${T_{PS}}$ and $k$) of above covariance matrixes. The selection probability of Bob and Alice only depend on their data, one-side non-Gaussian selection will not have impact on the other, which is also consistent with the above derivation.

Here, we assume Eve implements the correlated attacks~\cite{CVMDI2,CVMDI4,ATTACK} on two quantum channels. Then the modes ${C_x}$ and ${C_p}$ of Charlie are

\begin{equation}
\begin{aligned}
{C_x} &= \frac{1}{{\sqrt 2 }}\left( {{A_5} - {B_6}} \right)\\
&= \frac{1}{{\sqrt 2 }}\left( {\sqrt {{T_A}} {A_4} - \sqrt {{T_B}} {B_5}} \right)\\
& + \frac{1}{{\sqrt 2 }}\left( {\sqrt {1 - {T_A}} {E_2} - \sqrt {1 - {T_B}} {E_3}} \right),\\
{C_p} &= \frac{1}{{\sqrt 2 }}\left( {{A_5} + {B_6}} \right)\\
& = \frac{1}{{\sqrt 2 }}\left( {\sqrt {{T_A}} {A_4} + \sqrt {{T_B}} {B_5}} \right)\\
&  + \frac{1}{{\sqrt 2 }}\left( {\sqrt {1 - {T_A}} {E_2} + \sqrt {1 - {T_B}} {E_3}} \right).
\end{aligned}
\end{equation}
where $T_A$ ($T_B$) is the transmittance of the channel from Alice (Bob) to Charlie. Both channel loss of Alice and Bob are $\alpha  = 0.2{{{\rm{dB}}} \mathord{\left/
 {\vphantom {{{\rm{dB}}} {{\rm{km}}}}} \right.
 \kern-\nulldelimiterspace} {{\rm{km}}}}$, the transmittance of the channel are ${T_A} = {10^{ - \alpha {{{L_{AC}}} \mathord{\left/
 {\vphantom {{{L_{AC}}} {10}}} \right.
 \kern-\nulldelimiterspace} {10}}}}$ and ${T_B} = {10^{ - \alpha {{{L_{BC}}} \mathord{\left/
 {\vphantom {{{L_{BC}}} {10}}} \right.
 \kern-\nulldelimiterspace} {10}}}}$ respectively. After the displacement of Bob, the two quadratures of ${B_1}$ are

\begin{equation}
\begin{aligned}
{B_{1x}} &= {B_{2x}} + \mu {C_x} \\
&= \left( {{B_{2x}} - \mu \sqrt {\frac{{{T_B}}}{2}} {B_{5x}}} \right) + \mu \sqrt {\frac{{{T_A}}}{2}} {A_{4x}}\\
& + \frac{\mu }{{\sqrt 2 }}\left( {\sqrt {1 - {T_A}} {E_{2x}} - \sqrt {1 - {T_B}} {E_{3x}}} \right),\\
{B_{1p}} &= {B_{2p}} + \mu {C_p}\\
& = \left( {{B_{2p}} - \mu \sqrt {\frac{{{T_B}}}{2}} {B_{5p}}} \right) + \mu \sqrt {\frac{{{T_A}}}{2}} {A_{4p}}\\
& + \frac{\mu }{{\sqrt 2 }}\left( {\sqrt {1 - {T_A}} {E_{2p}} - \sqrt {1 - {T_B}} {E_{3p}}} \right).
\end{aligned}
\label{ce}
\end{equation}
It has been demonstrated that the CV-MDI QKD protocol could be seen as the one-way CV QKD protocol using coherent states and heterodyne detection by assuming that both Bob¡¯s source and the displacement operation inside himself are untrusted~\cite{CVMDI1}. Then according to Eqs.~(\ref{cmA}),~(\ref{cmB}) and~(\ref{ce}), the final covariance matrix  of $\rho_{A_1B_1}$ is given by

\begin{equation}
{\gamma _{{A_1}{B_1}}}  = \left[ {\begin{array}{*{20}{c}}
{{V_{{A_1}}}{\rm I}}&{\sqrt T {C_{{A_1}{A_4}}}{\sigma _z}}\\
{\sqrt T {C_{{A_1}{A_4}}}{\sigma _z}}&{\left[ {T\left( {{V_{{A_4}}} - 1} \right) + 1 + T\epsilon'} \right]{\rm I}}
\end{array}} \right],
\end{equation}
to simplify the equation, the transmittance and the excess noise are given by

\begin{equation}
\begin{aligned}
T &= \frac{{{T_A}}}{2}{\mu ^2}\\
\epsilon' &= 1 + \frac{1}{{{T_A}}}\left[ {{T_B}\left( {{\chi _B} - 1} \right) + {T_A}{\chi _A} - {C_E}} \right]\\
 &+ \frac{1}{{{T_A}}}\left[ {\frac{{\sqrt 2 }}{\mu }\sqrt {{V_{{B_5}}} - 1}  - \sqrt {{T_B}\left( {{V_{{B_2}}} + 1} \right)} } \right],
\end{aligned}
\label{CE}
\end{equation}
where ${C_E} = \frac{2}{{{T_A}}}\sqrt {\left( {1 - {T_A}} \right)\left( {1 - {T_B}} \right)} \left\langle {{E_{2x}}{E_{3x}}} \right\rangle $ (X quadrature) or ${C_E} = -\frac{2}{{{T_A}}}\sqrt {\left( {1 - {T_A}} \right)\left( {1 - {T_B}} \right)} \left\langle {{E_{2p}}{E_{3p}}} \right\rangle $ (P quadrature) is the noise contribution from the correlation of Eve's two modes.By setting $\mu  = \sqrt {\frac{{2\left( {{V_{{B_5}}} - 1} \right)}}{{{T_B}\left( {{V_{{B_2}}} + 1} \right)}}} $, the total excess noise is minimized as

\begin{equation}
\epsilon' = {\epsilon_A} + \frac{1}{{{T_A}}}\left[ {{T_B}\left( {{\epsilon_B} - 2} \right) + 2} \right].
\end{equation}

Combined with the above analysis, the secret key rate here is equivalent to the conventional one-way CV-QKD with coherent state and heterodyne detection. The increase in asymptotic rate also applies to non-asymptotic rate. For simplicity, the asymptotic secret key rate is calculated in this paper. The reverse reconciliation which has higher secret key rate than direct one is used in the following analysis. Thus the asymptotical secret key rate against collective attacks for reverse reconciliation~\cite{REVERSE} is derived as

\begin{equation}\label{kr}
{K_{PS}} = P\left[ {\beta I\left( {A:B} \right) - S(E:B)} \right],
\end{equation}
where $\beta $ is reconciliation efficiency, $P = P_{S}^{{A}} \cdot P_{S}^{{B}}$, $P_{S}^{{A}}$ ($P_{S}^{{B}}$) is the success probability of non-Gaussian post-selection with photon subtraction number $k$ in Alice's (Bob's) side, $I\left( {A:B} \right)$ is classical mutual information between Alice and Bob and $S(E:B) = S\left( E \right) - S\left( {E\left| B \right.} \right)$ is quantum mutual information between Eve and Bob.

The mutual information between Alice and Bob is calculated by

\begin{equation}
\begin{array}{l}
I\left( {A:B} \right) = \log \left( {{V_{{A_x}}}} \right) - \log \left( {{V_{{A_x}\left| {{B_x}} \right.}}} \right)\\
{\kern 1pt} {\kern 1pt} {\kern 1pt} {\kern 1pt} {\kern 1pt} {\kern 1pt} {\kern 1pt} {\kern 1pt} {\kern 1pt} {\kern 1pt} {\kern 1pt} {\kern 1pt} {\kern 1pt} {\kern 1pt} {\kern 1pt} {\kern 1pt} {\kern 1pt} {\kern 1pt} {\kern 1pt} {\kern 1pt} {\kern 1pt} {\kern 1pt} {\kern 1pt} {\kern 1pt} {\kern 1pt} {\kern 1pt} {\kern 1pt} {\kern 1pt} {\kern 1pt} {\kern 1pt} {\kern 1pt} {\kern 1pt} {\kern 1pt} {\kern 1pt} {\kern 1pt} {\kern 1pt} {\kern 1pt} {\kern 1pt} {\kern 1pt} = \log \left( {\frac{{{V_{{A_1}}} + 1}}{{{V_{{A_1}}} - \frac{{C_{{A_1}{B_1}}^2}}{{{V_{{B_1}}}}} + 1}}} \right),
\end{array}
\end{equation}
$V_{{A_1}}$, $C_{{A_1}{B_1}}$ and $V_{{B_1}}$ are corresponding elements of covariance matrix in Eq.~(\ref{pscm}). Eve can purify the whole system ${\rho _{{A_1}{B_1}}}$, so we can derive $S\left( E \right) = S({A_1}{B_1})$. $S({A_1}{B_1})$ which is a function of the symplectic eigenvalue ${\lambda _1}$ and ${\lambda _2}$ of ${\gamma _{{A_1}{B_1}}}$

\begin{equation}
S\left( {{A_1}{B_1}} \right) = \sum\limits_{i = 1}^2 {G\left( {{\lambda _i}} \right)}.
\end{equation}
where $G({\lambda _i}) = \frac{{{\lambda _i} + 1}}{2}{\log _2}\left( {\frac{{{\lambda _i} + 1}}{2}} \right) + \frac{{{\lambda _i} - 1}}{2}{\log _2}\left( {\frac{{{\lambda _i} - 1}}{2}} \right)$.

After Bob applies heterodyne detection on mode ${B_1}$, the system ${{A_1}E}$ is pure. Eve's condition entropy $S(E\left| {{x_{{B_1}}},{p_{{B_1}}}} \right.) = S({A_1}\left| {{x_{{B_1}}},{p_{{B_1}}}} \right.)$. $S({A_1}\left| {{x_{{B_1}}},{p_{{B_1}}}} \right.)$ is a function of the symplectic eigenvalue ${\lambda _3}$ of $\gamma _{{A_1}}^{{x_{{B_1}}},{p_{{B_1}}}}$

\begin{equation}
S({A_1}\left| {{x_{{B_1}}},{p_{{B_1}}}} \right.) = G\left( {{\lambda _3}} \right),
\end{equation}
and
\begin{equation}
\gamma _{{A_1}}^{{x_{{B_1}}},{p_{{B_1}}}} = {\gamma _{{A_1}}} - T{C_{{A_1}{B_1}}}{\left( {{\gamma _{{B_1}}} + {\rm I}} \right)^{ - 1}}C_{{A_1}{B_1}}^T.
\end{equation}

\section{\label{sec:3}Numerical simulation and analyses}

In CV-MDI QKD protocols, the performance of the asymmetric case (${T_A} \ne {T_B}$) has been proven to be superior to the symmetric case (${T_A}={T_B}$). It means that when Charlie's position is far away from Bob, the total transmission distance will drop rapidly to a few kilometers. When the distance between Charlie and Bob is 0, the transmittance distance of protocol will increase to the maximal distance with same parameters~\cite{CVMDI1}. Because the longer transmission distance is more suitable for point-to-point applications, the extreme asymmetric case is first utilized to simulate the secret key rate and tolerable excess noise. The secret key rate indicates the amount of the secret key distilled in one optical pulse and the tolerable excess noise indicates the amount of the excess noise which reduces the secret key rate to 0. According to the Eq.~\ref{CE}, the correlated attacks degenerate into two independent attacks in the extreme asymmetric scheme. By comparing these two crucial performances of CV-QKD, we can search the best scheme of three cases.

\section*{A. The performance in the extreme asymmetric scheme}

\begin{figure}[t]
\centering
\includegraphics[width=3.3in]{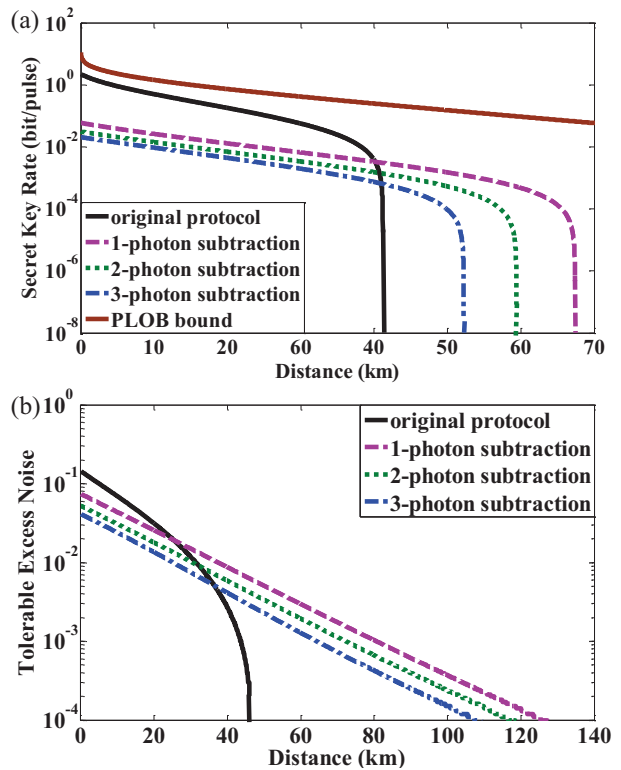}
\caption{ (Color online) (a) The maximal secret key rate with optimal $T_{PS}^A$ in (b) by setting different $k$ vs PLOB bound of equivalent photon subtraction only in Alice's side. (b) The maximal tolerable noise with optimal $T_{PS}^A$ by setting different $k$ of equivalent photon subtraction only in Alice's side. In the simulations, the variance of TMSV state is ${V_A} = {V_B} = 40$, the excess noise is $\epsilon = 0.002$, the reconciliation efficiency is $\beta  = 0.95$.
}\label{SAlice1}
\end{figure}

In CV-MDI QKD protocols, measurement results from Charlie will be used by Alice and Bob to generate secret key. Before the reconciliation, Bob uses ${x_B} = {x_{B1}} + \mu {x_{C}}$ and ${p_B} = {p_{B1}} - \mu {p_{C}}$ to modify his data. By this step, the information of Bob is almost removed from the measurement results. The remaining parts only include noise and part of Alice's information. As a result, the virtual photon subtraction applied by Alice with appropriate parameters may improve the performance of CV-MDI QKD.

The non-Gaussian post-selection success probability is determined by the equivalent numbers $k$ and transmittance ${T_{PS}}$ of virtual photon subtraction. We illustrate the effect of the non-Gaussian post-selection on the performance of the CV-MDI QKD protocol by the secret key rate and tolerable excess noise. We consider the performance with different parameters $k$ and the corresponding optimal $T_{PS}^A$ which is shown in Fig~\ref{SAliceAP}(b). The PLOB bound~\cite{PLOB} which is the secret-key capacity of the lossy channel indicates the maximum rate achievable by any optical implementation of QKD. The comparison of the performance between various 1,2,3-photon subtraction protocols and the PLOB bound is shown in Fig~\ref{SAlice1}(a). The numerical simulation shows that the secret key rate of the various 1,2,3-photon subtraction protocols is two orders lower than the PLOB bound. That means that the MDI node can not be an active repeater even in the improved CV-MDI protocol. In addition to the PLOB bound, the non-Gaussian post-selection with $k=1$ gets the furthest maximal transmittance distance (the secret key rate higher than 1e-8 is reserved) and the highest secret key rate at long transmittance distance. The tolerable noise with three $k$ is shown in Fig~\ref{SAlice1}(b), the renovated protocol with $k=1$ still performs better than other cases.

Although the secret key rate at long distance is enhanced, the short-distance performance is even worse than the original protocol with optimal $T_{PS}^A$. The first reason is the low success probability of the photon subtraction, which can not exceed $25\%$ under the parameters we used in this work. Another reason is that the photon subtraction will equivalently adjust the modulation variance. The optimal modulation variance is not a constant at different distances. To extend the transmittance distance, the photon subtraction is applied to optimize the modulation variance at long distance. As a result, the optimal modulation variance for long distance does not apply to short distance. Combining these two reasons, the secret key rate at short distance is almost two orders lower than the value at long distance. According to the above analysis, the virtual photon subtraction of Alice improves the performance of the renovated protocol at long transmittance distance where the number of photon subtraction is $k=1$ for the optimal choice of $T_{PS}^A$.

\begin{figure}[t]
\centering
\includegraphics[width=3.3in]{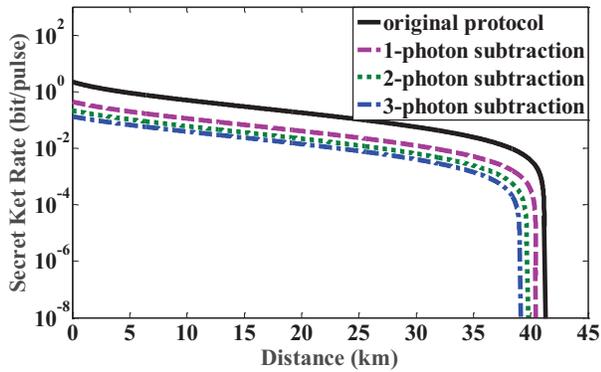}
\caption{ (Color online) The maximal secret key rate with optimal $T_{PS}^B$  by setting different $k$ of equivalent photon subtraction only in Bob's side. In the simulations, the variance of TMSV state is ${V_A} = {V_B} = 40$, the excess noise is $\epsilon = 0.002$, the reconciliation efficiency is $\beta  = 0.95$.
}\label{SBob1}
\end{figure}

\begin{figure}[t]
\centering
\includegraphics[width=3.3in]{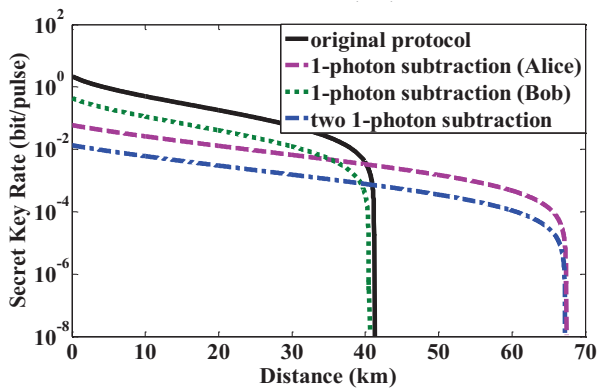}
\caption{ (Color online) The comparison of maximal secret key rate among the original CV-MDI QKD, the CV-MDI QKD with virtual photon subtraction in only Alice's side, Bob's side and both their side. In the simulations, the variance of TMSV state is ${V_A} = {V_B} = 40$, the excess noise is $\epsilon = 0.002$, the reconciliation efficiency is $\beta  = 0.95$.
}\label{SAliceBob}
\end{figure}

By assuming Bob's source and the displacement operation are untrusted, the covariance matrix of CV-MDI QKD is equivalent to the one-way form. The secret key rate which is calculated from the equivalent one-way form is independent on Bob's source. Thus, the virtual photon subtraction in Bob's side has little effect on the final secret key rate. On the other hand, Bob choose an optimal $\mu $ to make his final estimator of Alice's data more precise, but the virtual photon subtraction still reduces $T = \frac{{{T_A}}}{2}{\mu ^2}$ which is considered to be controlled by Eve and leads to a slight decline in performance. By the same way as above, we can get the optimal $T_{PS}^B$ with different $k$. In the case of the optimal $T_{PS}^B$, when the value of $k$ increase, the value of $T$ will drop accordingly. Thus when virtual photon subtraction is only in Bob's side, the secret key rate and tolerable excess noise will both be reduced. Moreover, the selection probability can not reach 100\%, which is also the cause of the performance decline. The secret key rate curves with different $k$ and optimal $T_{PS}^B$ are compared in Fig~\ref{SBob1}. The simulation results are consistent with the above theoretical analysis. As a result, compared to the renovated protocol, the original protocol can achieve higher secret key rate and tolerable excess noise. In summary, Bob's photon subtraction will not bring any benefit.

According to the results of above two simulations, the performance is improved when virtual photon subtraction is only in Alice's side at long distance. Because the two sources are independent, the impact of Alice and Bob's photon subtraction will independently act on the final secret key rate. Then we compare the performance of scheme with two 1-photon subtraction in each side to the scheme with virtual photon subtraction only in Alice or Bob's side. The comparison of secret key rate shown in Fig~\ref{SAliceBob} indicates that we cannot enhance the secret key rate higher by using two 1-photon subtraction in each side than the scheme with 1-photon subtraction only in Alice's side. The benefit of Alice's virtual photon subtraction is offset by the side effect of Bob's virtual photon subtraction. The simulation results show the improvement of Alice's photon subtraction and the damaging effect of Bob's photon subtraction. In summary, the performance of the modified protocol is only improved by Alice using virtual photon subtraction with reasonable parameters.

\section*{B. Symmetric scheme under the optimal correlated attacks}

\begin{figure}[t]
\centering
\includegraphics[width=3.3in]{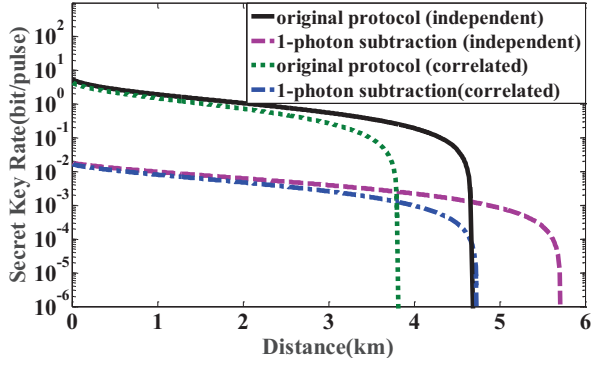}
\caption{ (Color online) The comparison of maximal secret key rate among the original CV-MDI QKD and the CV-MDI QKD with non-Gaussian post-selection under two independent attacks and the optimal correlated attacks. In the simulations, the variance of TMSV state is ${V_A} = {V_B} = 10000$, the excess noise is $\epsilon = 0.002$, the reconciliation efficiency is $\beta  = 0.95$.
}\label{SAliceCA}
\end{figure}

Although the performance of the symmetric CV-MDI QKD protocols is better at long distance, the symmetric case has potentials in short-range network applications where the relay should be in the middle of Alice and Bob. In the symmetric scheme, the correlated attacks are not equal to two independent attacks. The covariances of Eve's two modes could be any value, which should satisfy the uncertainty principle ~\cite{GATTACK}. The optimal correlated attacks are proven to be the negative ERP attack where the ancillas are maximally entangled and negatively related~\cite{CVMDI4}. The performance of the original protocols and the renovated protocols under two kinds of attacks are compared in the Fig~\ref{SAliceCA}.

Because CV-MDI protocols have been showed to be vulnerable to correlated attacks on the link, the secret key rate of the original protocol and renovated protocol under the optimal attacks are both lower than the key rate under two independent attacks case. No matter under either attacks, the transmittance distance is still enhanced by the virtual photon subtraction. As a result, the virtual photon subtraction is also effective to improve the long-distance performance of CV-MDI protocols under the optimal attacks.
\section{\label{sec:4}Conclusion}
The CV-MDI QKD protocols, where the secret key between two legal parties is established by the measurement results of an untrusted third party, are impossible to improve the performance at the detection side. We propose a method to improve the performance of coherent-state CV-MDI QKD protocols by non-Gaussian post-selection of transmitted data. Non-Gaussian post-selection is a practical way to apply ideal photon subtraction which can equivalently enhance the entanglement of TMSV state. We prove that the virtual photon subtraction of Alice and Bob's transmitted data are independent of each other. For CV-MDI QKD with virtual photon subtraction used by Bob, the performance is decreased due to the lower equivalent channel transmittance and the limited success probability. The parameters of non-Gaussian post-selection which is used by Alice can be adjust flexibly to improve the secret key rate and tolerable excess noise at long transmission distance. In summary, we provide an efficient optimization method for long-range application of CV-MDI QKD.

Note added. Shortly after the submission of our manuscript, an independent work~\cite{MDIPS} has been posted on arXiv. This work studied the performance of the CV-MDI QKD protocol with photon subtraction only used by Alice.

\section*{Acknowledgments}

This work was supported by the Key Program of National Natural Science Foundation of China under Grant 61531003, the National Natural Science Foundation under Grant 61427813 and 61501414, the National Basic Research Program of China (973 Pro-gram) under Grant 2014CB340102 and the Fund of State Key Laboratory of Information Photonics and Optical Communications.

\section*{APPENDIX A: Calculation of the covariance matrix}

\begin{figure}[t]
\centering
\includegraphics[width=3.3in]{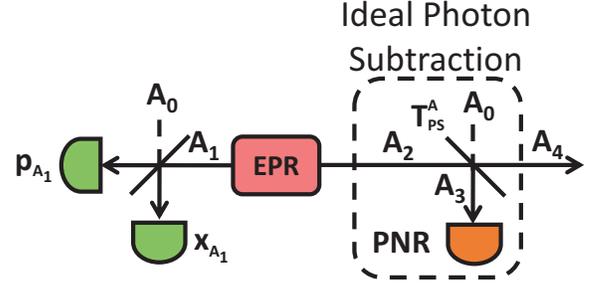}
\caption{ (Color online) Entanglement-based scheme of photon-subtracted TMSV source using an ideal photon number resolving detector. ${T_{PS}}$ is transmittance of the beam splitter. PNR is the photon number resolving detector.}
\label{TMSV}
\end{figure}

A TMSV state is generated by squeezing a two mode vacuum state ${\left| \psi  \right\rangle _{{A_1}{A_2}}}$. The two mode squeezing operator is ${S_{TMS}}\left( r \right) = \exp \left[ {{{r\left( {{{\hat a}_1}{{\hat a}_2} - \hat a_1^\dag \hat a_2^\dag } \right)} \mathord{\left/
 {\vphantom {{r\left( {{{\hat a}_1}{{\hat a}_2} - \hat a_1^\dag \hat a_2^\dag } \right)} 2}} \right.
 \kern-\nulldelimiterspace} 2}} \right]$, where ${\hat a_1}$ and $\hat a_1^\dag $ ( ${\hat a_2}$ and $\hat a_2^\dag $ ) represent the annihilation and creation operator of modes ${A_1}$ and ${A_2}$, $r$ is the squeezing parameter. Thus the TMSV state ${\left| \psi  \right\rangle _{{A_1}{A_2}}}$  can be described as

\begin{equation}
\begin{aligned}
{\left| \psi  \right\rangle _{{A_1}{A_2}}} &= {S_{TMS}}\left( r \right)\left| {0,0} \right\rangle\\
  &= \sqrt {1 - {\lambda ^2}} \sum\limits_{n = 0}^\infty  {{\lambda ^n}\left| {n,n} \right\rangle } ,
\end{aligned}
\end{equation}
where $\lambda  = \tanh r$, $\left| {n,n} \right\rangle  = {\left| n \right\rangle _{{A_1}}} \otimes {\left| n \right\rangle _{{A_2}}}$ and $\left| n \right\rangle $ is Fock state.

Photon subtraction is implemented by a beam splitter (transmittance $T$) and a photon number resolving detector shown in Fig.~\ref{TMSV}. The mode ${A_2}$ of the TMSV state is split into ${A_3}$ and ${A_4}$ by the beam splitter. Then the TMSV state ${\left| \psi  \right\rangle _{{A_1}{A_2}}}$ is transformed into a tripartite state ${\left| \psi  \right\rangle _{{A_1}{A_3}{A_4}}}$

\begin{equation}
{\left| \psi  \right\rangle _{{A_1}{A_2}{A_3}}} = {U_{BS}}\left( T \right)\left| {TMSV} \right\rangle \left| 0 \right\rangle,
\end{equation}

The photon number resolving detector is described by a Positive Operator-Valued Measurement (POVM) operators $M_{PS}^k = \left| k \right\rangle \left\langle k \right|$. ${\left| \psi  \right\rangle _{{A_1}{A_2}{A_3}}}$ with k photons subtraction can be written as

\begin{equation}
\left| \psi  \right\rangle _{{A_1}{A_2}{A_3}}^k = \frac{{M_{PS}^k{{\left| \psi  \right\rangle }_{{A_1}{A_2}{A_3}}}}}{{\sqrt {{{\left\langle \psi  \right|}_{{A_1}{A_2}{A_3}}}{M_{ps}}{{\left| \psi  \right\rangle }_{{A_1}{A_2}{A_3}}}} }},
\end{equation}
where $\sqrt {{{\left\langle \psi  \right|}_{{A_1}{A_2}{A_3}}}{M_{ps}}{{\left| \psi  \right\rangle }_{{A_1}{A_2}{A_3}}}}  = P_{PS}^k$ is success probability of subtracting $k$ photons also the probability of detecting $k$ photons on mode ${A_3}$ on the photon number resolving detector.

When Alice replaces the original source by the photon subtracted TMSV state, she first prepares a TMSV state and applies the heterodyne detection on mode ${A_1}$. There is a one-to-one correspondence between preparing a coherent state and measuring one mode of the TMSV state. The mode ${A_2}$ is separated into two modes ${A_3}$ and ${A_4}$ by a beam splitter and mode ${A_3}$ is measured by photon subtraction $M_{PS}^k$. The mode with k-photon subtraction is described by

\begin{figure}[t]
\centering
\includegraphics[width=3.3in]{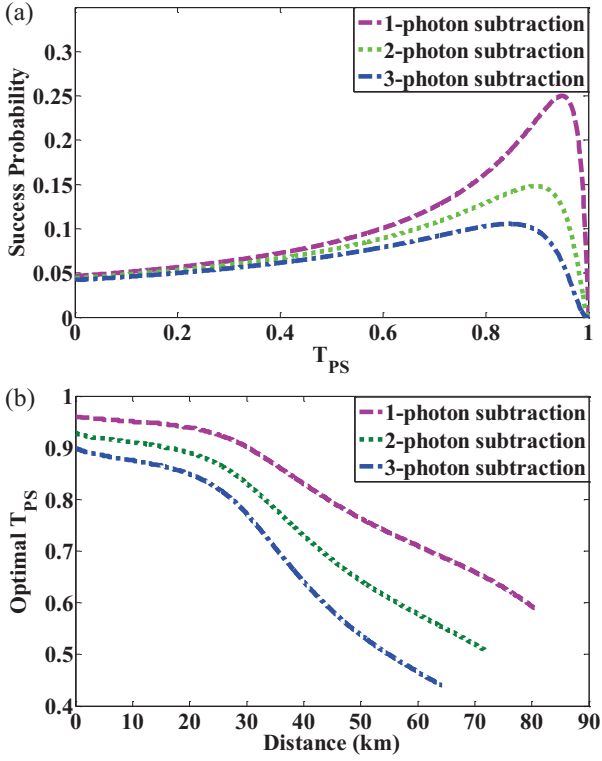}
\caption{ (Color online) (a) The success probability $P_{S}^k$ of virtual photon subtraction by setting different transmittance ${T_{PS}}$. (b) The optimal $T_{PS}^A$ of virtual photon subtraction for the maximal secret key rate by setting different $k$. In the simulations, the variance of TMSV state is ${V_A} = {V_B} = 40$, the excess noise is $\epsilon = 0.002$ and the reconciliation efficiency is $\beta  = 0.95$.}
\label{SAliceAP}
\end{figure}

\begin{equation}
\rho _{{A_4}}^k = {P_{{A_{1x}}{A_{1p}}}}d{A_{1x}}d{A_{1p}}\left| {\sqrt {{T_{PS}}} \alpha } \right\rangle \left\langle {\sqrt {{T_{PS}}} \alpha } \right|,
\end{equation}
where $\alpha  = \sqrt 2 \lambda ({A_{1x}} + i{A_{1p}})$, ${P_{{A_{1x}}{A_{1p}}}}$ is the two-dimension Gaussian distribution of Alice's measurement result $\left( {{A_{1x}},{A_{1p}}} \right)$ and the success probability of k-photon subtraction with Alice's heterodyne measurement result is $P_{PS}^k\left( {k\left| {{A_{1x}},{A_{1p}}} \right.} \right)$

\begin{equation}
\begin{aligned}
P_{PS}^k\left( {k\left| {{A_{1x}},{A_{1p}}} \right.} \right) = &\exp \left[ { - \frac{{\left( {1 - {T_{PS}}} \right){\lambda ^2}}}{2}\left( {A_{1x}^2 + A_{1p}^2} \right)} \right]\\
&{{ \cdot {{\left[ {\frac{{\left( {1 - {T_{PS}}} \right){\lambda ^2}}}{2}\left( {A_{1x}^2 + A_{1p}^2} \right)} \right]}^k}} \mathord{\left/
 {\vphantom {{ \cdot {{\left[ {\frac{{\left( {1 - {T_{PS}}} \right){\lambda ^2}}}{2}\left( {A_{1x}^2 + A_{1p}^2} \right)} \right]}^k}} {k!}}} \right.
 \kern-\nulldelimiterspace} {k!}},
\end{aligned}
\end{equation}

Then the covariance matrix of photon subtracted TMSV state can be described by

\begin{equation}
{\gamma _{{A_1}{A_4}}} = \left[ {\begin{array}{*{20}{c}}
{{V_{{A_1}}}{\rm I}}&{{C_{{A_1}{A_4}}}{\sigma _z}}\\
{{C_{{A_1}{A_4}}}{\sigma _z}}&{{V_{{A_4}}}{\rm I}}
\end{array}} \right],
\end{equation}
Because of the symmetry of covariance matrix, we can get the matrix element by only calculating the variance and covariance of quadrature ${A_{4x}}$

\begin{equation}
\begin{array}{l}
{V_{{A_1}}} = 2\int {\int {A_{1x}^2P\left( {{A_{1x}},{A_{1p}},{A_{4x}}} \right)d{A_{1x}}d{A_{1p}}d{A_{4x}} - 1} } \\
{C_{{A_1}{A_4}}} = \sqrt 2 \int {\int {{A_{1x}}{A_{4x}}P\left( {{A_{1x}},{A_{1p}},{A_{4x}}} \right)d{A_{1x}}d{A_{1p}}d{A_{4x}}} } \\
{V_{{A_4}}} = \int {\int {A_{4x}^2P\left( {{A_{1x}},{A_{1p}},{A_{4x}}} \right)d{A_{1x}}d{A_{1p}}d{A_{4x}}} },
\end{array}
\end{equation}
where
\begin{equation}
\begin{aligned}
P\left( {{A_{1x}},{A_{1p}},{A_{4x}}} \right) =& \frac{{P_{PS}^k\left( {k\left| {{A_{1x}},{A_{1p}}} \right.} \right)}}{{P_{PS}^k}}\\
 &\cdot {P_{{A_{1x}}{A_{1p}}}}{\left| {\left\langle {{A_{4x}}} \right.\left| {\sqrt {{T_{PS}}} \alpha } \right\rangle } \right|^2},
\end{aligned}
\end{equation}

and
\begin{equation}
P_{PS}^k = \frac{{1 - {\lambda ^2}}}{{1 - {T_{PS}}{\lambda ^2}}}{\left[ {\frac{{{\lambda ^2}\left( {1 - {T_{PS}}} \right)}}{{1 - {T_{PS}}{\lambda ^2}}}} \right]^k}.
\end{equation}

\section*{APPENDIX B:  The optimal transmittance of photon subtraction's beam splitter}

The success probability of virtual photon subtraction $P_{S}^{{A}}$ is displayed as a function of transmittance $T_{PS}^A$ in Fig.~\ref{SAliceAP}(a) when the variance of original TMSV state is constant. As a result, the transmittance $T_{PS}^A$ has effect on the secret key rate ${K_S} = \beta I\left( {B:A} \right) - S(E:B)$ without consideration of success probability and the success probability of non-Gaussian post-selection simultaneously. The final secret key rate is dependent on both above parts. Thus, there is a tradeoff between the secret key rate ${K_S}$ and the success probability $P_{S}^{{A}}$ to maximize the final secret key rate. By comparing the secret key rate with different transmittance $T_{PS}^A$, the optimal transmittance $T_{PS}^A$ with differen $k$ is illustrated in Fig~\ref{SAliceAP}(b).

\end{document}